\documentclass{ws-procs975x65}            % default, citations in superscript
\newcommand{\be}{\begin{equation}}
\newcommand{\ee}{\end{equation}}

\begin{document}
\title{$H_0$ from cosmic chronometers and Type Ia supernovae, with Gaussian processes and the weighted polynomial regression method}

\author{Adri\`a G\'omez-Valent\footnote{Speaker}}

\address{Departament de F\'isica Qu\`antica i Astrof\'isica, and Institute of Cosmos Sciences,\\ Universitat de Barcelona, Av. Diagonal 647, E-08028 Barcelona, Catalonia, Spain \\
adriagova@fqa.ub.edu}

\author{Luca Amendola}
\address{Institut f\"{u}r Theoretische Physik, Ruprecht-Karls-Universit\"{a}t Heidelberg\\ Philosophenweg 16, 69120 Heidelberg, Germany\\l.amendola@thphys.uni-heidelberg.de}

\begin{abstract}
We derive new constraints on the Hubble parameter $H_0$ using the available data on $H(z)$ from cosmic chronometers (CCH), and the Hubble rate data points from the supernovae of Type Ia (SnIa) of the Pantheon compilation and the Hubble Space Telescope (HST) CANDELS and CLASH Multy-Cycle Treasury (MCT) programs. We employ two alternative techniques, Gaussian Processes (GPs) and the Weighted Polynomial Regression (WPR) method, to reconstruct the Hubble function, determine the derived values of $H_0$, and compare them with the local HST measurement provided by Riess {\it et al.} (2018), $H_0^{{\rm HST}}=(73.48\pm1.66)$ km/s/Mpc, and with the Planck+$\Lambda$CDM value, $H_0^{{\rm P18}}=(66.88\pm 0.92)$ km/s/Mpc. With GPs we obtain $H_0=(67.99\pm 1.94)$ km/s/Mpc, and with the WPR method $H_0=(68.90\pm 1.96)$ km/s/Mpc. Both are fully compatible at $<1\sigma$ c.l., and also with $H_0^{{\rm P18}}$. In contrast, they are in $\sim 2\sigma$ tension with $H_0^{{\rm HST}}$.
\end{abstract}


\bodymatter

%%%%%%%%%%%%%%%%%%%%%%%%%%%%%%%%%%%%%%%%%%%%%%%%%%%%%%%%%%%%%%%%%%
%%%%%%%%%%%%%%%%%%%%%%%%%%%%%%%%%%%%%%%%%%%%%%%%%%%%%%%%%%%%%%%%%%

\section{Introduction}\label{introduction1}
The Hubble-Lema\^itre constant, $H_0$, is a crucial parameter in Cosmology. It does not only tell us about the current velocity of recession of distant astronomical objects due to the cosmological dynamics, but it is also essential to compute cosmic distances and spans of time. Now, ninety years after the discovery of the Universe's expansion and first measurement of $H_0$ by Edwin Hubble \cite{Hubble1929}, and after the many works in the literature since then aiming to improve the estimations of this parameter, we can state with high statistical confidence level that $H_0$ lies in the range of $65-75$ km/s/Mpc. Nevertheless, this parameter is still a matter of many discussions between the members of the cosmological community, mainly focused on the origin of the huge tension between the local (distance ladder) determination of $H_0$ obtained with the Hubble Space Telescope (HST)\cite{RiessH02018} and the value inferred from the temperature and polarization anisotropies of the cosmic microwave background (CMB) radiation measured by the Planck satellite\cite{Planck2018}, by assuming the $\Lambda$CDM theoretical framework. The former reads $H^{{\rm HST}}_0=(73.48\pm 1.66)$ km/s/Mpc, whereas the TT+lowE CMB Planck+$\Lambda$CDM analysis leads to $H_0^{{\rm P18}}=(66.88\pm 0.92)$ km/s/Mpc. These values are in $\sim 3.5\sigma$ tension. Finding the ultimate reason that generates this mismatch has become one of the most important issues in Cosmology. In principle, unaccounted systematics affecting the HST and/or Planck measurements could play a role here. Regarding this point, it is important to mention that many other works making use of the cosmic distance ladder find values of $H_0$ also lying in the high region preferred by Ref. \refcite{RiessH02018}, see e.g. Refs. \refcite{Cardona2017,Zhang2017,JangLee2017}. There are also discrepant voices in the literature, though. For instance, the author of Ref. \refcite{Romano2018} argues that the use by Riess et al. of the 2M++ density field map (which covers redshifts $z\leq 0.06$) to compute peculiar velocity flows could be biasing their results, since there is evidence of the existence of local radial inhomogeneities extending in different directions up to a redshift of about $0.07$ \cite{Keenan2013}, and according to Ref. \refcite{Romano2018} the $40\%$ of the Cepheids used in Ref. \refcite{RiessH02018} would be affected. Moreover, the authors of Ref. \refcite{Shanks2018a} claim that the GAIA parallax distances of Milky Way Cepheids employed in Ref. \refcite{RiessH02018} in the first step of the cosmic distance ladder may be underestimated a $\sim 7-18\%$. This would produce an important decrease of their measured value of $H_0$. The discussion on the validity of these arguments is, though, still open and intense.\cite{RiessProbShanks2018b} 

In regards to the potential systematics affecting Planck's data, it is worth to highlight the study of Ref. \refcite{Addison2017}, in which the authors showed that the use of CMB data independent from Planck together with data from baryon acoustic oscillations (BAO) and the big bang nucleosynthesis (BBN) gives rise to values of $H_0$ fully compatible with the Planck preferred range and, hence, it seems quite improvable that systematics affect Planck's analysis in a decisive way.

Assuming the $\Lambda$CDM, the authors of Refs. \refcite{Bonvin2017,Birrer2018} analyzed the gravitational time delay of the light rays coming from multiply imaged quasar systems and found values of $H_0$ unable to discriminate between the two values in dispute due to their large error bars. It has also been possible to measure the Hubble parameter using the gravitational wave signal of the neutron star merger GW17081716 and its electromagnetic counterpart \cite{Abbott2017,Guidorzi2017,Hotokezaka2018}, but still with very big uncertainties too. This method is interesting because it is cosmology-independent and represents an alternative to the cosmic distance ladder measurement\cite{RiessH02018}. According to Ref. \refcite{Feeney2018}, a sample of $\sim 50$ binary neutron star standard sirens (detectable within the next decade) will be able to arbitrate between the local and CMB estimates. The impact of the cosmic variance on the local determination of $H_0$ has been also studied in Refs. \refcite{Marra2013,WuHuterer,CamarenaMarra}. The authors conclude that its effect cannot explain the whole discrepancy between the HST and Planck's values.

In view of all these facts and the unsuccessful efforts of the community of finding a theoretical model able to relieve in an efficient way the $H_0$-tension (see e.g. Refs. \refcite{Bernal2016,PLB2017}), it is important to go on with the investigations on the value of the Hubble parameter. Particularly relevant are those studies that try to extract values from observations at intermediate redshifts, as model-independent as possible, and also utilizing approaches different from the one applied in Ref. \refcite{RiessH02018}, where the authors made use of the cosmic distance ladder and low-redshift SnIa ($z<0.15$). We present here the main results of our dedicated work \cite{GomezAmendola2018}, in which we employed data on cosmic chronometers (CCH) and supernovae of Type Ia (SnIa), and two different methods -- Gaussian Processes (GPs) and the novel Weighted Polynomial Regression (WPR) technique -- to reconstruct the Hubble function and derive an extrapolated value of $H_0$.  
 
%%%%%%%%%%%%%%%%%%%%%%%%%%%%%%%%%%%%%%%%%%%%%%%%%%%%%%%%%%%%%%%%%%
%%%%%%%%%%%%%%%%%%%%%%%%%%%%%%%%%%%%%%%%%%%%%%%%%%%%%%%%%%%%%%%%%%

\section{Data}

In this section we limit ourselves to cite the references from which we have collected the data on CCH and SnIa, and confer the reader to Ref. \refcite{GomezAmendola2018} and references therein for a more detailed account and description of these data sets. On the one hand, we employ the first 5 (correlated) effective points on the normalized Hubble rate, i.e. $E(z)=H(z)/H_0$, from the Pantheon+MCT sample \cite{Scolnic2018,Riess2018b}, which includes 1063 SnIa. As it is shown in Ref. \refcite{Riess2018b}, the compression of the information contained in such SnIa sample is carried out in a very efficient way. It is well-known that SnIa cannot be used alone to determine $H_0$, since this parameter is fully degenerated with the SnIa absolute magnitude. To break such degeneracy, we use the 31 (uncorrelated) Hubble function data points from cosmic chronometers provided in Refs. \refcite{Jimenez2003,Simon,Stern,MorescoA,Zhang,MorescoB,MorescoC,Ratsismbazafy2017}. These data cover a redshift range up to $z\sim 2$ and are obtained without assuming any particular cosmological model, by applying spectroscopic dating techniques of passively-evolving galaxies. They do not rely neither on the Cepheid distance scale nor parallaxes. There are, though, other sources of systematic uncertainties, as the ones associated to the modeling of stellar ages, which is carried out making use of the so-called stellar population synthesis (SPS) models, e.g. the BC03\cite{BC03} and MaStro\cite{MaStro} ones. Let us detail which SPS models have been used in obtaining the CCH data points of Refs. \refcite{Jimenez2003,Simon,Stern,MorescoA,Zhang,MorescoB,MorescoC,Ratsismbazafy2017}. In Refs. \refcite{Zhang,Stern,Ratsismbazafy2017} the authors only provide the values of $H(z_i)$ obtained with the BC03 model. This constitutes a $\sim 25\%$ of the whole CCH data set. In Ref. \refcite{MorescoB} only the combined MaStro/BC03 values are available, whereas in Refs. \refcite{Jimenez2003,Simon} an alternative SPS model is used, different from the MaStro and BC03 ones. These points constitute the $\sim 32\%$ of the CCH data set. In contrast, in Refs. \refcite{MorescoA,MorescoC} the authors provide both, the BC03 and MaStro values. This group includes the remaining $\sim 43\%$ of the data. We have opted to use the BC03 values of these two references in our main analyses so as to incorporate consistently the data from Refs. \refcite{Zhang,Stern,Ratsismbazafy2017}, namely to avoid the use of a mixture of MaStro and BC03 values while maximizing the number of data points entering the calculations. The impact of considering alternative data sets has been discussed in detail in Sect. 3.4 of Ref. \refcite{GomezAmendola2018}. We will also comment later on on ({\it sic}) the effect of using a more conservative data set which takes into account the combination of BC03 and MaStro values of Refs. \refcite{MorescoA,MorescoC} instead of only the BC03 ones. This data set coincides with the one provided in Table 2 of Ref. \refcite{qjGomezValent}. We refer the reader to this paper for details.

%%%%%%%%%%%%%%%%%%%%%%%%%%%%%%%%%%%%%%%%%%%%%%%%%%%%%%%%%%%%%%%%%%
%%%%%%%%%%%%%%%%%%%%%%%%%%%%%%%%%%%%%%%%%%%%%%%%%%%%%%%%%%%%%%%%%%

\section{Reconstructing $H(z)$ from observations}

%%%%%%%%%%%%%%%%%%%%%%%%%%%%%%%%%%%%%%%%%%%%%%%%%%%%%%%%%%%%%%%%%%
%%%%%%%%%%%%%%%%%%%%%%%%%%%%%%%%%%%%%%%%%%%%%%%%%%%%%%%%%%%%%%%%%%

\subsection{With Gaussian Processes}
Gaussian distributions are defined for a finite set of quantities, and are characterized by a vector of mean values and the corresponding covariance matrix, which is in charge of controlling the uncertainties' size and the correlations. Gaussian Processes are their direct generalization to the continuum, and are analogously characterized by a mean function $\mu(z)$ and a two-point covariance function $\mathcal{C}(z,z^\prime)$, see e.g. Ref. \refcite{GPbook},
\begin{equation}\label{eq:GP1}
\xi(z)\sim \mathcal{GP} \left(\mu(z),\mathcal{C}(z,z^\prime)\right)\,,
\end{equation}
where the curve $\xi(z)$ is a realization of the Gaussian process. The covariance $\mathcal{C}(z,z^\prime)$ is defined as follows: (i) when $z$ and/or $z^\prime$ do {\it not} coincide with any point contained in the data set that we want to use to reconstruct the function, then $\mathcal{C}(z,z^\prime)=\mathcal{K}(z,z^\prime)$, where $\mathcal{K}(z,z^\prime)$ is the so-called kernel function, which must be of course symmetric and at this stage is unknown; (ii) when both, $z$ and $z^\prime$, {\it do} coincide with the $z$-values of one or more points contained in our data set, then $\mathcal{C}(z,z^\prime)=\mathcal{D}(z,z^\prime)+\mathcal{K}(z,z^\prime)$. Here we incorporate the information of the known covariance matrix $\mathcal{D}(z,z^\prime)$ of our data points. The kernel function plays a crucial role in the GPs, and must be selected beforehand. Three of the most famous ones with only two degrees of freedom are the following: (a) The Gaussian kernel, $\mathcal{K}(z,z^\prime)=\sigma_f^2{\rm Exp}[-\frac{1}{2}\left(\frac{z-z^\prime}{l_f}\right)^2]$; (b) The Cauchy kernel, $\mathcal{K}(z,z^\prime)=\sigma_f^2l_f/[(z-z^\prime)^2+l_f^2]$; and (c) The Mat\'ern kernel, $\mathcal{K}(z,z^\prime)=\sigma_f^2\left[1+\frac{\sqrt{3}}{l_f}|z-z^\prime|\right]{\rm Exp}[-\frac{\sqrt{3}}{l_f}|z-z^\prime|]$. They all depend on $\sigma_f$ and $l_f$, the so-called hyperparameters of the kernel function. The first one controls the uncertainties' size and the strength of the correlations, whereas the second somehow limits the scope of these correlations in $z$. Although it is possible to construct more elaborated kernels, e.g. kernels which do not only depend on the distance $|z-z^\prime|$, but also on the locations $z$ and $z^\prime$ themselves, in such a way that the symmetry under the interchange $z\leftrightarrow z^\prime$ is kept intact, we will stick to the more simple ones mentioned above and, more concretely, only to the Gaussian one, since as we showed in Ref. \refcite{GomezAmendola2018} the reconstructed $H(z)$ obtained with these three kernels are compatible at $<1\sigma$ c.l. for all the redshift range under study. Of course, the results also depend on the values of the hyperparameters. Once we choose the kernel, how can we properly select $\sigma_f$ and $l_f$? To do so we must make use of our data. In the GPs philosophy, our data set is conceived as part of a subset of realizations of the Gaussian process. The hyperparameters are usually chosen so as to maximize the probability of the GP to produce our data set. If we marginalize the GP \eqref{eq:GP1} over the points at $z^*$, i.e. those at which we do not have data, we get the following multivariate normal distribution, 
\begin{equation}\label{eq:GP2}
\vec{\xi}\sim \mathcal{N} \left(\{\mu_i(\tilde{z}_i)\},\mathcal{C}\right)\,,
\end{equation}
where $i=1,...,N$, with $N$ being the dimension of the vector of data points $\vec{y}\equiv\{\tilde{z}_i,y_i\}$ at our disposal, and $\mu_i(\tilde{z}_i)$ can be set e.g. to $\vec{0}$ $\forall{i}$, since the result is almost insensitive to this. Thus, the hyperparameters will be obtained upon the minimization of 
\begin{equation}\label{eq:logL}
-2\ln\mathcal{L}(\sigma_f,l_f)=N\ln(2\pi)+\ln|\mathcal{C}(\sigma_f,l_f)|+\vec{y}^T\mathcal{C}^{-1}(\sigma_f,l_f)\vec{y}\,,
\end{equation}
with $\mathcal{L}$ being the marginal likelihood and $|\mathcal{C}|$ the determinant of $\mathcal{C}$. Having done this, and using \eqref{eq:GP1}, we can compute the conditional probability of finding a given realization of the Gaussian process in the case in which $\xi(\tilde{z}_i)=y_i(\tilde{z}_i)$. The resulting mean and variance functions extracted from such conditioned GP read, respectively,
\begin{equation}
\bar{\xi}(z^*)=\sum_{i,j=1}^{N}\mathcal{C}^{-1}(\tilde{z}_i,\tilde{z}_j)y(\tilde{z}_j)\mathcal{K}(\tilde{z}_i,z^*)\,,
\end{equation}
\begin{equation}
\sigma^{2}(z^*)=\mathcal{K}(z^*,z^*)-\sum_{i,j=1}^{N}\mathcal{C}^{-1}(\tilde{z}_i,\tilde{z}_j)\mathcal{K}(\tilde{z}_i,z^*)\mathcal{K}(\tilde{z}_j,z^*)\,.
\end{equation}
\begin{figure}[t]
\centering
\includegraphics[scale=0.5]{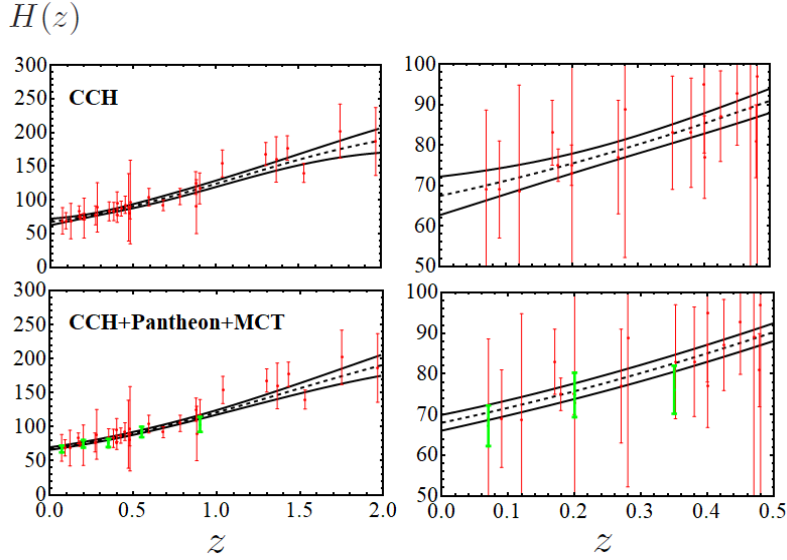}
\caption{Reconstructed $H(z)$ in [km/s/Mpc] with the corresponding $1\sigma$ bands obtained with GPs when only CCH are used (figures on the top), and when we add the Pantheon+MCT data (the ones at the bottom). In both cases we zoom in the redshift range $z\in[0,0.5]$ (plots on the right) so as to better appreciate our determination for $H_0$ and how the uncertainty bands narrow down when the SnIa data are included. The processed ($E(z_i)\rightarrow H(z_i)$) Pantheon+MCT points are plotted in green, whereas the CCH data are in red. See related comments in the main text.}
\end{figure}
The results derived using this formalism are shown in Fig. 1. When only CCH are included in the analysis we obtain an extrapolated value of $H_0=(67.42\pm 4.75)$ km/s/Mpc. This result coincides with the one provided in Ref. \refcite{YuRatraWang2017}. We have incorporated the SnIa information with the following procedure. We have firstly used the value of $H_0$ extracted from the GP-reconstruction of $H(z)$ with CCH to promote (via a Monte Carlo routine) the Hubble rates of Ref. \refcite{Riess2018b} to values on $H(z_i)$. They can be used now in combination with the CCH data in a new GP-reconstruction to obtain the corresponding new value of $H_0$. The result reads, $H_0=67.99\pm 1.94$ km/s/Mpc. The error bars has decreased a factor $\sim 2.5$ with respect to when only CCH are employed. The central value is fully compatible with the Planck+$\Lambda$CDM result, and it is in $2.2\sigma$ tension with $H_0^{\rm HST}$. Using the more conservative CCH data set of Ref. \refcite{qjGomezValent} in combination with the Pantheon+MCT data one obtains $H_0=69.00\pm 2.35$ km/s/Mpc. In this case the tension with the HST value loosens, being now of $1.7\sigma$. For a discussion on the effect of the propagation of the hyperparameters' uncertainties we refer the reader to Sect. 3.3 of Ref. \refcite{GomezAmendola2018}.

%%%%%%%%%%%%%%%%%%%%%%%%%%%%%%%%%%%%%%%%%%%%%%%%%%%%%%%%%%%%%%%%%%
%%%%%%%%%%%%%%%%%%%%%%%%%%%%%%%%%%%%%%%%%%%%%%%%%%%%%%%%%%%%%%%%%%

\subsection{With the Weighted Polynomial Regression method}
Cosmographical analyses are based on truncated expansions (in $z$ or some other variable) of cosmological quantities, e.g. cosmic distances or the Hubble function, around $z=0$, see e.g. Ref. \refcite{Visser}. People usually truncate these expansions (which do not rely on any particular cosmological model) at a concrete order and subsequently fit the resulting expression to the data in order to extract kinematic information of our Universe. The truncation order should not be determined in an {\it ad hoc} way, but applying some well-motivated criterion, e.g. looking for the order that makes the value of the reduced chi-squared statistic to be as close to one as possible, the one that maximizes the Bayesian evidence (see e.g. Ref. \refcite{DEbook}), or just the one that minimizes some approximate information criterion statistic, as the Akaike \cite{Akaike} or Bayesian \cite{BIC} ones. But even these approaches can prove insufficient to obtain objective (non-biased) constraints on the fitting parameters, as it has been highlighted in Refs. \refcite{GomezAmendola2018,qjGomezValent}. However, it is possible to improve the usual cosmographical methodology by making use of the so-called Weighted Function Regression technique, which does not consist on choosing only one particular order of the truncated series, but on weighting in a consistent way the contribution of the various truncated expansions. 

Let us explain how it works. Firstly, imagine that we have the following function, which is linear in the parameters,
\begin{equation}\label{eq:fitFunc}
f(z)=\sum_{i=0}^{n}a_ib_i(z)\,,
\end{equation}
with the $b_i$'s being the so-called basis functions and $\vec{a}$ the vector of coefficients. If we have the vector of mean values $\vec{\bar{a}}$ and the associated covariance matrix $D$, then we can compute the mean function $\bar{f}_M(z)$ and covariance matrix ${\rm cov}[f_M(z),f_M(z^\prime)]$ as follows: 
\begin{equation}\label{eq:meanRecFunc}
\bar{f}_M(z)=\sum_{i=0}^{n}\bar{a}_ib_i(z)\,,\qquad {\rm cov}[f_M(z),f_M(z^\prime)]=\sum_{i,j=0}^{n} D_{ij}b_i(z)b_j(z^\prime)\,.
\end{equation}
$\vec{\bar{a}}$ and $D$ can be computed exactly in the case of Gaussian-distributed data with flat priors (cf. Ref. \refcite{GomezAmendola2018}) or, in a more general case, computationally making use of a Monte Carlo routine. The variance of the reconstructed function is just $\sigma_M^2(z)={\rm cov}[f_M(z),f_M(z)]$. These tools can be used, e.g. to reconstruct $H(z)$ in the cosmographic scenario. We just have to Taylor-expand $H(z)$ around e.g. $z=0$,
\begin{equation}\label{eq:TaylorH}
H(z)=H_0+\frac{dH}{dz}\Bigr\vert_{z=0}z+\frac{1}{2}\frac{d^2H}{dz^2}\Bigr\vert_{z=0}z^2+...
\end{equation}
In this case $b_i(z)=z^i$, $a_0=H_0$, $a_1=H_0(q_0+1)$, etc. Let us call $M_0$, $M_1$,..., $M_{N-1}$ the cosmographic polynomials of order $n=0$, $1$,..., $N-1$, respectively, with $N$ being again the number of data points entering the analysis. That is, let us conceive each polynomial as a different model, and compute the probability density associated to the fact of having a certain shape of the function $f(z)$ as follows,
\begin{equation}
P[f(z)]=k\cdot[P(f(z)|M_0)P(M_0)+...+P(f(z)|M_{N-1})P(M_{N-1})]\,,
\end{equation}
where $k$ is just a normalization constant that must be fixed by imposing $\int[\mathcal{D}f]\,P[f(z)]=1$. Taking into account that $\int[\mathcal{D}f]\,P(f(z)|M_J)=1\quad \forall J\in[0,N-1]$ and $\sum_{J=0}^{N-1} P(M_J)=1$, we find $k=1$ and therefore:
\begin{equation}
P[f(z)]=\sum_{J=0}^{N-1}P(f(z)|M_J)P(M_J)\,.
\end{equation}
We now denote $M_*$ as the most probable model and identify $\frac{P(M_J)}{P(M_*)}$ with the Bayes ratio $B_{J*}$ \cite{DEbook}. The last expression can be finally written as
\begin{equation}\label{eq:finExp}
P[f(z)]=\frac{\sum\limits_{J=0}^{N-1}P(f(z)|M_J)B_{J*}}{\sum\limits_{J=0}^{N-1}B_{J*}}\,.
\end{equation}
This is the central expression of the WPR method, where the weights are directly given by the Bayes factors. Notice that using \eqref{eq:finExp} we can compute the (weighted) moments and related quantities too, e.g. the weighted mean and variance read,
\begin{equation}\label{eq:MeanTotal}
\bar{f}(z)=\frac{\sum\limits_{J=0}^{N-1}\bar{f}_{J}(z)B_{J*}}{\sum\limits_{J=0}^{N-1}B_{J*}}\,,\qquad \sigma^2(z)=\frac{\sum\limits_{J=0}^{N-1}[\sigma_J^2(z)+(\bar{f}_{J}(z))^2]B_{J*}}{\sum\limits_{J=0}^{N-1}B_{J*}}-(\bar{f}(z))^2\,,
\end{equation}
where $\bar{f}_{J}(z)$ and $\sigma^2_J(z)$ can be computed by means of \eqref{eq:meanRecFunc}. For the obtention of the Bayes factors we invoke the time-honored Bayesian and Akaike information criteria, BIC and AIC, defined as \cite{Akaike,BIC}:
\begin{equation}\label{eq:criteria}
{\rm AIC}_J=\chi^2_{{\rm min},J}+\frac{2n_JN}{N-n_J-1}\,,\qquad{\rm BIC}_J=\chi^2_{{\rm min},J}+n_J\ln N\,,
\end{equation}
with $\chi^2_{{\rm min},J}$ being the minimum of the $\chi^2$ function in the model $M_J$, and $n_J=J+1$ the dimension of $\vec{a}_J$. In this way the Bayes factor between the model $M_J$ and the most probable model $M_*$ can be approximated by 
\begin{equation}\label{eq:Bayesratio}
B_{J*}=e^{\frac{BIC_*-BIC_J}{2}}\qquad {\rm or}\qquad B_{J*}=e^{\frac{AIC_*-AIC_J}{2}}\,,
\end{equation}
%  
%%%%%%%%%%%%%%%%%%%%%%%%%%%%%%%%
%%%%%%%%%%%%%%%%%%%%%%%%%%%%%%%%

\begin{table}[t!]
\tbl{Fitting results obtained for the models with largest weight entering the WPR-reconstruction of $H(z)$ with the CCH data}
{\begin{tabular}{@{}ccccccc@{}}
\toprule
{\small Polyn. degree (d)} & $\chi^2_{{\rm min}}$ & BIC & AIC & $H_0$ & $B_{{\rm d}1}$ (BIC)  & $B_{{\rm d}1}$ (AIC) \\ \hline
 $1$& $16.62$ & $23.48$  & $21.04$ & $62.3\pm 3.1$ & $1$ & $1$\\ \hline
 $2$& $14.74$ & $25.04$ & $21.63$ & $67.8\pm5.1$  & $0.47$ & $0.74$\\  \hline
 $3$& $14.78$ & $28.51$ & $24.31$ & $68.9\pm4.5$ &  $0.08$ & $0.19$\\ \hline
 $4$& $14.80$ & $31.97$ & $27.20$ & $68.2\pm4.2$ & $0.014$ & $0.05$\\ \hline
 $5$& $14.95$ & $35.56$ & $30.45$ & $68.4\pm4.0$ &  $0.002$ & $0.009$ \\ \hline
 $6$& $15.01$ & $39.05$ & $33.88$ & $68.6\pm4.4$ &  $4\times 10^{-4}$ & $0.002$\\ \hline
 $7$& $15.27$ & $42.74$ & $37.82$ & $68.3\pm4.2$ &  $6\times 10^{-5}$ & $2\times 10^{-4}$\\ \hline
 $8$& $15.72$ & $46.63$ & $42.29$ & $68.8\pm4.4$ &  $9\times 10^{-6}$ & $2\times 10^{-5}$\\ \hline
 $9$& $16.38$ & $50.72$ & $47.38$ & $68.9\pm3.9$ &  $1\times 10^{-6}$ & $2\times 10^{-6}$\\ \hline
\end{tabular}}
\begin{tabnote}
Some quantities obtained from the fitting analysis of the first nine (cosmographic) polynomials involved in the WPR-reconstruction, by only using the CCH data. In the second column we show the minimum value of the $\chi^2$ function. The third and fourth columns contain the values of the Bayesian and Akaike information criteria, as defined in \eqref{eq:criteria}. In the fifth column we show the values of the Hubble parameter in [km/s/Mpc] together with its $1\sigma$ uncertainty. The last two columns contain the values of the Bayes ratio that are obtained by using the BIC and AIC, respectively, as defined in \eqref{eq:Bayesratio}. See comments in the text.\\
\end{tabnote}
\label{aba:table1}
\end{table}
%%%%%%%%%%%%%%%%%%%%%%%%%%%%%%%%
\noindent depending on the criterion used, the BIC or AIC, respectively. The most probable model $M_*$ is defined as the one with lowest BIC or AIC, depending again on the chosen criterion. It is crystal-clear from these expressions that the competing models with more parameters used to analyze the same data receive a suitable penalty and, therefore, our weighted method implements in practice Occam's razor principle. 

We apply the WPR formalism to the family of models for $H(z)$ that result from truncating the Taylor series \eqref{eq:TaylorH} at different orders. We force the coefficients to be positive so as to fulfill: (i) $H_0\ne 0$; (ii) $H^\prime(z=0)\ne 0$; and (ii) $dH/dz(z)>0\,\forall z>0$, cf. Ref. \refcite{GomezAmendola2018} for further details. We present in Table 1 the most relevant quantities obtained from the individual fitting analyses of the truncated expansions entering the WPR-reconstruction with CCH data. Therein we show the values of $H_0$ and the associated uncertainties derived from the individual fits and also the relative weight with respect to the most probable result, which in this case is obtained for the linear polynomial. Applying \eqref{eq:MeanTotal} we are led to $H_0=(64.44\pm 4.72)$ km/s/Mpc [$H_0=(65.27\pm 4.98)$ km/s/Mpc] when BIC [AIC] is used. When we add the Pantheon+MCT data we obtain $H_0=(68.90\pm 1.96)$ km/s/Mpc regardless of the information criteria. If we use the conservative CCH data set from Ref. \refcite{qjGomezValent} we obtain $H_0=(70.50\pm 2.60)$ km/s/Mpc. These results are completely compatible with the ones obtained using GPs.

%%%%%%%%%%%%%%%%%%%%%%%%%%%%%%%%%%%%%%%%%%%%%%%%%%%%%%%%%%%%%%%%%%
%%%%%%%%%%%%%%%%%%%%%%%%%%%%%%%%%%%%%%%%%%%%%%%%%%%%%%%%%%%%%%%%%%

\section{Conclusions}
In this work we have presented some estimations of $H_0$ obtained with two alternative reconstruction techniques and data on cosmic chronometers and supernovae of Type Ia. The results are independent from those obtained in Ref. \refcite{RiessH02018}, in which the authors made use of low-redshift SnIa and the cosmic distance ladder. Our results do not rely on any particular cosmological model, and seem to favor the low estimates of $H_0$ provided by the Planck Collaboration\cite{Planck2018}. Our results are also resonant with the conclusions of Ref. \refcite{Lin2017}, where the authors provided some indication that the local measurement \cite{RiessH02018} of the Hubble-Lema\^itre constant is an outlier, by using the so-called index of inconsistency to test the consistency between the preferred values of $H_0$ derived from alternative data sources: CMB from Planck, SnIa+BAO+BBN, data on large scale structure formation, gravitational time delay, and local determination of $H_0$. Other authors \cite{YuRatraWang2017,Feeney2018,Haridasu,Macaulay} have also obtained values of $H_0$ lying in the lower range preferred by Planck, using model-independent techniques and different combinations of intermediate-redshift data from BAO, SnIa, and CCH. Moreover, several analyses have shown that it is possible to loosen or even solve the $\sigma_8$-tension at the expense of keeping the one concerning $H_0$, see e.g. Refs. \refcite{PLB2017,AGVsola,Anand2017}. A low value of $H_0$ would automatically allow to remove  the existing tension in the $\sigma_8$-parameter.

Our methodology has proved to be a viable route for estimating $H_0$. Revisiting it in the future, when new and more precise data are available, would be interesting, especially when the theoretical uncertainties associated to the SPS models employed to extract the CCH data from observations decreases in a significant way.

%%%%%%%%%%%%%%%%%%%%%%%%%%%%%%%%%%%%%%%%%%%%%%%%%%%%%%%%%%%%%%%%%%
%%%%%%%%%%%%%%%%%%%%%%%%%%%%%%%%%%%%%%%%%%%%%%%%%%%%%%%%%%%%%%%%%%

\section*{Acknowledgements}
AGV would like to thank Prof. Joan Sol\`a for his invitation to give this talk as part of the CM4 parallel sessions on ``Tensions on $\Lambda$CDM cosmological model and model-independent constraints'' at the 15th Marcel Grossmann Meeting, in the beautiful city of Rome. He is also grateful to the organizers and the {\it Department of Quantum Physics and Astrophysics} of the University of Barcelona for their economical support. LA is supported by the DFG through TRR33 ``The Dark Universe''.

\end{document}